\documentclass{aa}  

\usepackage{graphicx}
\usepackage{txfonts}
\usepackage{multirow}
\usepackage{natbib}
\begin{document}

   \title{GeMS MCAO observations of the Galactic globular cluster NGC~2808: the absolute age}

   \subtitle{}

   \author{D. Massari\inst{1,2}
          \and
          G. Fiorentino\inst{1}
          \and
          A. McConnachie\inst{3}
          \and
          G. Bono\inst{4,5}
          \and
          M. Dall'Ora\inst{6}
          \and
          I. Ferraro\inst{5}
          \and
          G. Iannicola\inst{5}
          \and
          P.B. Stetson\inst{3}
          \and
          P. Turri\inst{7}
          \and
          E. Tolstoy\inst{2}      
          }

   \institute{INAF-Osservatorio Astronomico di Bologna, via Ranzani 1, 40127, Bologna, Italy\\
              \email{davide.massari@oabo.inaf.it}
         \and
            University of Groningen, Kapteyn Astronomical Institute, NL-9747 AD Groningen, Netherlands
         \and   
            Herzberg Astronomy and Astrophysics, National Research Council Canada, 5071 West Saanich Road, Victoria, BC V9E 2E7, Canada
         \and
            Dipartimento di Fisica, Universit\`{a} di Roma Tor Vergata, Via della Ricerca Scientifca 1, 00133 Roma, Italy
         \and
            INAF-Osservatorio Astronomico di Roma, Via Frascati 33, 00040 Monteporzio Catone, Italy
         \and
            INAF–Osservatorio Astronomico di Capodimonte, Via Moiariello 16, 80131 Napoli, Italy
         \and   
            Department of Physics and Astronomy, University of Victoria, 3800 Finnerty Road, Victoria, BC V8P 5C2, Canada
             }

   \date{Received November 3, 2015; accepted December 9, 2015}

 
  \abstract
   {Globular clusters are the oldest stellar systems in the Milky Way and probe the early epoch
   of the Galaxy formation. However, the uncertainties on their absolute age are still too large
   to soundly constrain how the Galactic structures have assembled.}
   {The aim of this work is to obtain an accurate estimate of the absolute age of the globular cluster NGC~2808 using
   deep IR data obtained with the multi conjugate adaptive optics system operating at the Gemini South telescope (GeMS).}
   {This exquisite photometry, combined with that obtained in V and I bands with HST, allowed us the detection of the faint 
   Main Sequence Knee feature in NGC~2808 colour magnitude diagram. 
   The difference between this point and the main sequence turn off is a good age estimator and provides 
   ages with unprecedented accuracy.}
   {We found that NGC~2808 has an age of t$=10.9\pm0.7$ (intrinsic) $\pm0.45$ (metallicity term) Gyr. A possible contamination by He-enhanced 
   population could make the cluster up to $0.25$ Gyr older. Although this age estimate agrees with the age coming from the classical
   turn off method (t$=11.0$ Gyr), its uncertainty is a factor $\sim3$ better, since it avoids systematics in reddening, distance assumptions and
   photometric zero points determination.
   The final absolute age indicates that NGC~2808 is slightly younger than other Galactic globular clusters with similar metallicity.}
   {}


   \maketitle
%

\section{Introduction}

The future advent of extremely large telescopes (ELTs), such as the Giant Magellan Telescope (GMT, see e.g. \citealt{thomas10}),
the Thirty Meter Telescope (TMT, \citealt{sanders13}) and the European-Extremely Large Telescope (E-ELT, \citealt{gilmozzi08}) is putting
near-infrared (NIR, $0.8$-$2.2$ $\mu$m) astronomy on the cutting edge of the next scientific revolution.
With diameters of their primary mirrors at least a factor of three larger than the 
current biggest telescopes, distant and unexplored regions of the universe will be
reached for the first time. At the same time, the potentially superb
spatial resolution will allow us to resolve extremely dense regions such as the centre
of our Galaxy and to study the astrometry of celestial bodies with unprecedented accuracy.
However, in order to fully exploit this potential, diffraction limited observations
have to be achieved and this requires the successful use and development of adaptive optics
(AO) techniques.

Up to now, the best AO-performance in terms of stability
of the AO-correction over a wide field of view (FoV) is reached by 
Multi-Conjugate AO (MCAO) systems such as the Multi-Conjugate Adaptive Optics
Demonstrator MAD (\citealt{marchetti07}) at the Very Large Telescope (see e.g. the 
works by \citealt{ferraro09a, bono10c, fiorentino11}) or the Gemini Multi-Conjugate Adaptive 
Optics System (GeMS, see e.g. \citealt{neichel14a, neichel14b, davidge14, saracino15, turri}) facility.
The latter is currently the only operating MCAO facility. 

GeMS uses 3 Natural Guide Stars (NGS) and five sodium Laser Guide Stars (LGS) to correct
for the distortions in the wave front by the thermally turbulent atmosphere.
A similar combination of LGS+NGS is planned for the E-ELT MCAO facility MAORY 
(Multi-conjugate Adaptive Optics RelaY for the E-ELT, \citealt{diolaiti10}). 
Therefore, the investigation of current GeMS 
performance and the exploitation of its data provide us a unique opportunity to 
learn how to take fully advantage of the future ELT potential.

Ideal scientific targets to study the MCAO performance are stellar fields ensuring a large
number of point-like sources uniformly distributed across 
the field of view, i.e. Galactic globular clusters (GCs). 
Our group have selected a sample of GCs and observed them with GeMS
with the aim of studying their near-IR properties, kinematics, and to understand the 
performance of an MCAO system using science based metrics, with an eye towards future 
science programs with the ELTs. First results from this program are presented in \cite{turri}.
Among these GCs, one of the most interesting is the southern GC NGC~2808, since it is
one of the most peculiar stellar systems in the Galaxy. With a mass of
$\sim10^{6}$ M$\odot$ (\citealt{mclaughlin05}) it is one of the most massive known GCs.
Its colour magnitude diagrams (CMDs) reveal several peculiar
features. \cite{dantona05} firstly discovered that the main sequence (MS) of
the cluster has a blueward extension not compatible with a single stellar population.
By using the {\it Hubble Space Telescope} Advanced Camera for Survey (HST/ACS),
\cite{piotto07} discovered that the MS of the cluster shows a complex structure
with the presence of three separated sequences. Such a complexity has further
increased now, since \cite{milone15} found the presence of at least five different
MSs and red giant branches (RGBs) by using a combination of ultra-violet (UV) HST filters.
\cite{bellini15} discovered two out of these five populations to have a radially anisotropic
velocity distribution.
All these findings, together with the investigation of other CMD sequences like
the peculiar Horizontal Branch (HB, see \citealt{bedin00, dalessandro11b, milone14}) or the RGB bump 
(\citealt{nataf13}) further support the hypothesis that multiple stellar 
generations in NGC~2808 are enriched in Helium by at least $\Delta Y=0.13$.
Such evidence has been confirmed spectroscopically by the work of \cite{bragaglia10} 
and from the direct measurement of the He coronal line at 10830~\AA by \cite{pasquini11}, 
thus demonstrating that NGC~2808 shows one of the most extreme Y-enhancement observed among all the Galactic GCs.
Moreover, the complexity of the system in terms of five multiple populations has
been confirmed spectroscopically by the recent work of \cite{carretta15}, who
also demonstrated its homogeneity in iron content.

Another peculiarity of NGC~2808 is its age. 
According to several works such as \cite{marin09} or \cite{vdb13}, this
cluster is substantially ($10-20$\%) younger than the average age of
the GCs with similar metallicity, being $\sim11$ Gyr old. However, such a claim does not
find a complete agreement in the literature, since for instance \cite{deangeli05}
estimated a younger age of $8.4$ Gyr using the difference in magnitude between the
MS-turn off (MSTO) and the Zero-Age HB (ZAHB), while \cite{piotto07}
found their CMD to be best fitted by a 12.5 Gyr old theoretical model.
By exploiting deep IR MAD observations, \cite{bono10c} defined a new method
to measure the absolute age of stellar clusters that is more precise than 
the previous ones by at least a factor of two. This method is based on
the magnitude difference $\Delta_{MSTO}^{MSK}$ between the MSTO and the faint
MS-knee (MSK), a feature which arises in the IR CMD due to an opacity effect 
driven by the collisionally induced absorption (CIA) of molecular
hydrogen (\citealt{saumon94, borysow01}) in the surface of cool dwarfs. 
Such a feature has also been observed in optical bands by \cite{dicecco15}
in the cluster M71, thus confirming the prediction of \cite{borysow97}.
MCAO IR observations are ideal to probe the faint
GC-MS, and in this paper we exploit GeMS observations of NGC~2808
to provide an accurate age measurement using this
newly developed method. 

The paper is organised as follows. In Section \ref{data} we present
our dataset and the details of the data reduction. In Section \ref{cmd}
the outputs of the reduction in terms of observed CMD are shown.
In Section \ref{results} we describe the result of our analysis
and we finally draw our conclusion in Section \ref{concl}.


\section{Photometric dataset and data reduction}\label{data}

This work is part of a project using Gemini-S (Program IDs: GS-2012B-SV-406, GS-2013A-Q-16, GS-2013B-Q-55, PI: McConnachie) to 
study the NIR photometric properties of Galactic GCs through the 
exploitation of the GeMS facility. 
GeMS assists the Gemini South Adaptive Optics Imager (GSAOI) IR Camera, which
is equipped with a $2\times2$ mosaic of Rockwell HAWAII-2RG $2048\times2048$
pixels array and covers a total FoV of $85$\arcsec $\times 85$\arcsec
with a spatial resolution of $0.02$ \arcsec pixel$^{-1}$ (\citealt{neichel14a}).
This MCAO system uses three natural guide stars and a constellation of five laser guide
stars to correct for the distortions introduced by the turbulent atmosphere of the Earth.

The dataset we analysed samples NGC~2808 central regions in the J and Ks filters,
through a series of 8 dithered exposures per filter, each with an exposure time of 160 seconds.
To overcome saturation of the brightest stars in these deep images,
we also used two 90 seconds-long and two 21 seconds-long short exposures per filter (see
Table~\ref{datatab}).

The pre-reduction of each raw image has been performed by means of the standard IRAF\footnote{IRAF is 
distributed by the National Optical Astronomy Observatory, which
is operated by the Association of Universities for Research in Astronomy, Inc.,
under cooperative agreement with the National Science Foundation} tools.
The master flat-field has been built by combining dome-flat fields (which have high signal to noise
ratio) and twilight-flat fields (which have a uniform illumination). We retrieved 20 dome-flat fields 
taken with projector lamps turned off (OFF), 20 taken with lamps turned on (ON) and 20 twilight-flat fields (TWI).
We computed the 3 $\sigma$-clipped median frame for each of the three, and defined the 
final master flat (FL) as FL=(ON$_{med}$-OFF$_{med}$)$\times$R, where R is the ratio between the
normalized TWI$_{med}$ and (ON$_{med}$-OFF$_{med}$), which is further smoothed with a median filter
to suppress the pixel-to-pixel statistics and leave only a term that describes the illumination
pattern of the sky. In this way the master flat has the pixel-to-pixel statistics of the dome-flats
and the illumination pattern of the sky-flats.
We did not need to apply dark frames because of the very low dark current of the H2RG detectors.

\begin{figure*}
    \includegraphics[width=\columnwidth]{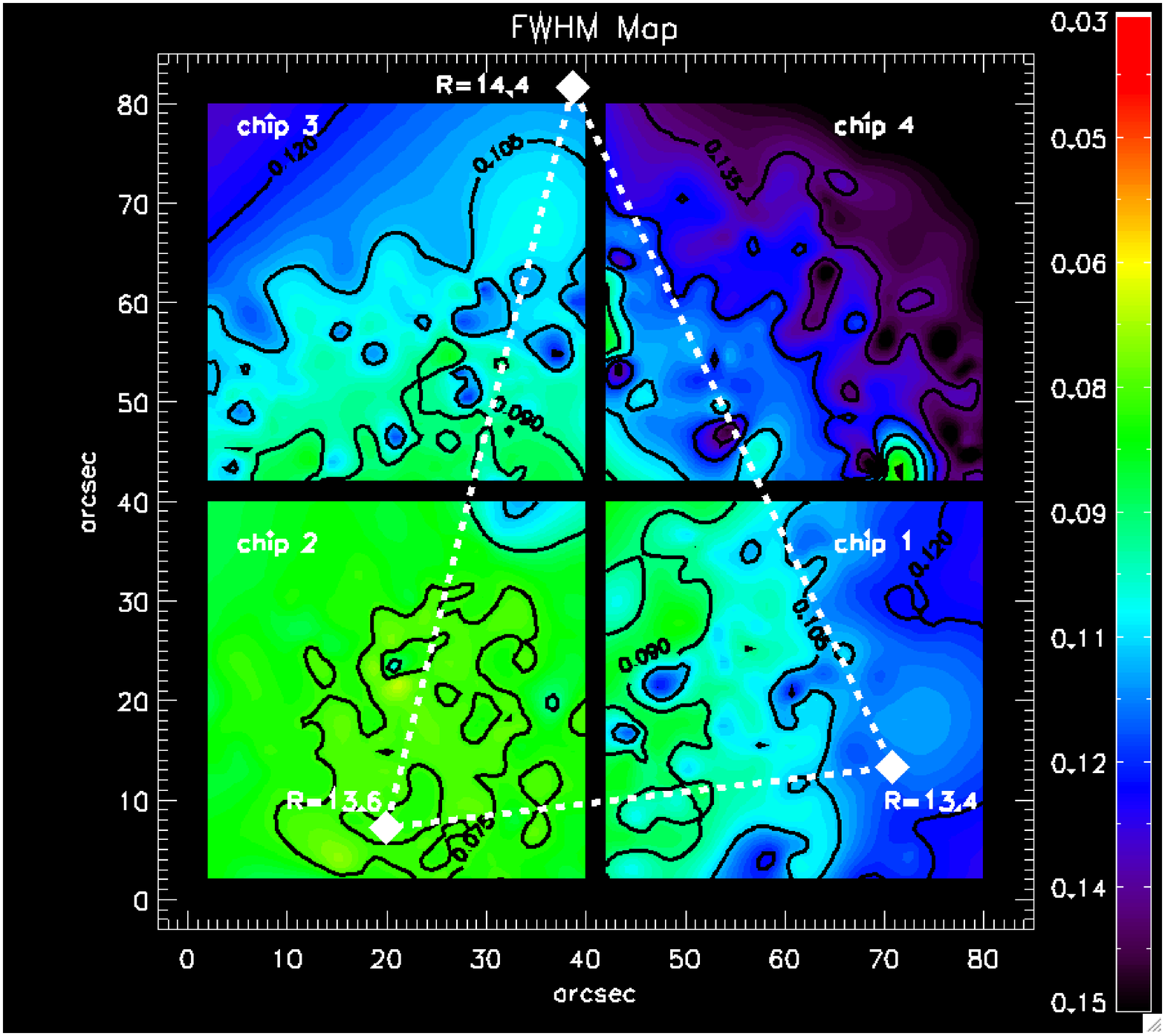}
    \includegraphics[width=\columnwidth]{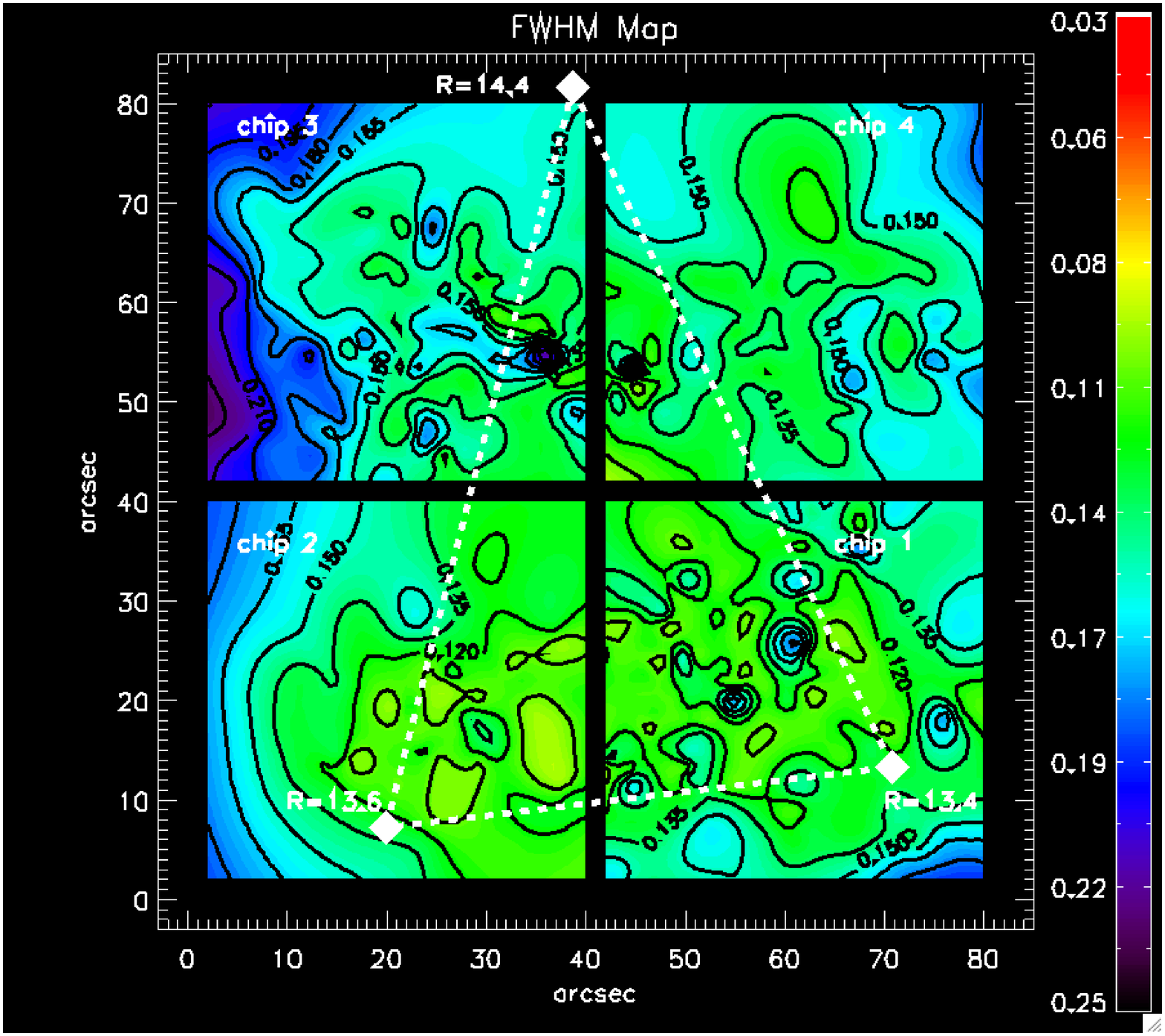}
        \caption{\small FWHM map for stars measured in one of the Ks-band (left panel) and J-band (right panel) images. The former was taken on the 19th of April 2013,
        the latter on the 23rd of May 2103. Chips are numbered counter-clockwise
        starting from the lower-right chip 1 (see white labels). The white squares indicate the location
        of the three Natural Guide Stars. Iso-FWHM contours are also overplotted as thick black lines.}\label{fwhm}
\end{figure*}

Unfortunately, the observing conditions were not very good.
As summarised in Table~\ref{datatab}, the mean seeing\footnote{Seeing values are taken from the Cerro Pachon seeing monitor (DIMM).
We are aware that the DIMM usually gives higher values than the true seeing measured on scientific images (see e.g. \citealt{fiorentino11}). However,
it is the only proxy we can use to quantitatively describe the quality of the weather.} during the first night (19th of April 2013) 
was $\sim0.6$\arcsec, and only 2 short- and 6 long-exposures in the Ks-band were taken. During the subsequent 
two nights when the observations were completed (April the 20th and May the 23rd, 2013), the seeing 
always exceeded 1.1\arcsec. Left panel of Fig.\ref{fwhm} shows the AO-corrected full width at half maximum (FWHM) variation 
across the GeMS FoV for one of the K-band images. Chips are numbered from 1 to 4 starting from the lower-right
chip, moving clockwise. From this map it is evident that the better 
and more stable correction has been achieved in the lower-left chip number 2, but that outside
the triangle defined by the location of the Natural Guide Star (NGS), the AO-correction is not as effective.
Across the entire FoV, the K-band FWHM varies from $0.07$\arcsec to $0.15$\arcsec (diffraction limit is $\sim0.06$\arcsec), 
that is a total variation of about the 50\%, which is a worse performance than found e.g. by \cite{neichel14b, saracino15}.
In this sense it is worth mentioning that one of the NGS (located in the northern region of the image)
is considerably fainter (R$\simeq14.4$ mag) 
than the other two, being $\sim 1$ magnitude less luminous in the R-band. Such an asymmetry might be responsible
for the worse photometric performance obtained for the two upper chips.

\begin{table}
\caption{GeMS photometric dataset}             
\label{datatab}      
\centering                          
\begin{tabular}{c c c c}        
\hline\hline                 
Night & Filter & t$_{exp}$ (s) & Seeing (\arcsec) \\    
\hline                        
    \multirow{3}{*}{April 19th} &  & $2\times21$ & 0.5 \\
     & K & $2\times90$ & 0.6\\
     &  & $6\times160$ & 0.7\\ 
\hline       
    \multirow{1}{*}{April 20th} & K & $2\times160$ & 1.2 \\
\hline
    \multirow{3}{*}{May 23rd} &  & $2\times21$ & 1.2 \\
     & J & $2\times90$ & 1.2\\
     &  & $8\times160$ & 1.2\\
\hline     
\end{tabular}
\end{table}

The right panel of Fig.\ref{fwhm} shows the same map for one of the J-band images. The combination of
intrinsically less efficient AO-correction at these shorter wavelengths and the poor seeing 
conditions means the situation is even worse. Here the FWHM varies from $\sim0.07$\arcsec in a
small region of chip 2 to $\sim0.24$\arcsec in the left side of chip 3, with a typical value
across the entire FoV of $\sim0.15$\arcsec (diffraction limit is $\sim0.04$\arcsec). 
As we will show in Sect.\ref{cmd}, this strongly affects the depth achieved by our IR photometry.

The reduction of the scientific images has been performed by means of the DAOPHOT suite of software
(\citealt{stetson87, stetson94}). The chips of each image have been treated separately.
The PSF for each frame has been modelled with great care, by selecting $\sim100-200$ bright and 
isolated stars uniformly distributed across the FoV and by fitting their brightness profile
with an analytic function plus a look-up table of the residuals that has been allowed to vary
spatially as a cubic function. The best-fit model has then been applied to all the sources
detected above a 3$\sigma$ threshold from the local background by using ALLSTAR.
In order to improve both the astrometry and the photometry on each single image, we then
created for each filter and chip a master list of stars composed of all the sources detected
in at least one image and provided it as an input for ALLFRAME.
The output files have been combined with DAOMASTER and 4 single-chip
catalogues with J and K instrumental magnitudes have been built by selecting only stars detected in 
at least 4 (2 K-band and 2 J-band) images and averaging their position and magnitudes measurements.

\begin{figure}
    \includegraphics[width=\columnwidth]{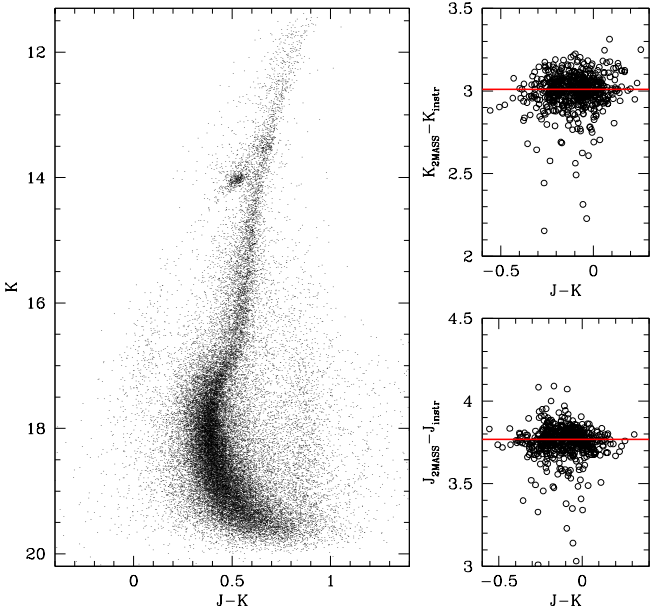}
        \caption{\small {\it Left panel:} K, J-K CMD of NGC~2808 from HAWK-I images. Instrumental magnitudes
        have been calibrated onto the 2MASS reference system by using the relations shown in the {\it right panels}, using
        $495$ stars in common between the two catalogues.}\label{cal_hawki}
\end{figure}

Each catalogue has then been calibrated onto the Two Micron All Sky Survey (2MASS) photometric scale by
means of an intermediate catalogue of calibrated stars obtained through the analysis of 25
High Acuity Wide field K-band Imager (HAWK-I, see \citealt{kissler08}) exposures of the same cluster
in the J and K filters.
The reduction of these images followed the same steps described before. By cross-matching our
HAWK-I catalogue with the 2MASS catalogue, we found $495$ stars in common and we used them
as calibrators. The final calibrated HAWK-I (K,J-K) colour magnitude diagram (CMD) is shown in the left
panel of Fig. \ref{cal_hawki}, together with the relations used to calibrate HAWK-I instrumental
magnitudes.

\begin{figure}
    \includegraphics[width=\columnwidth]{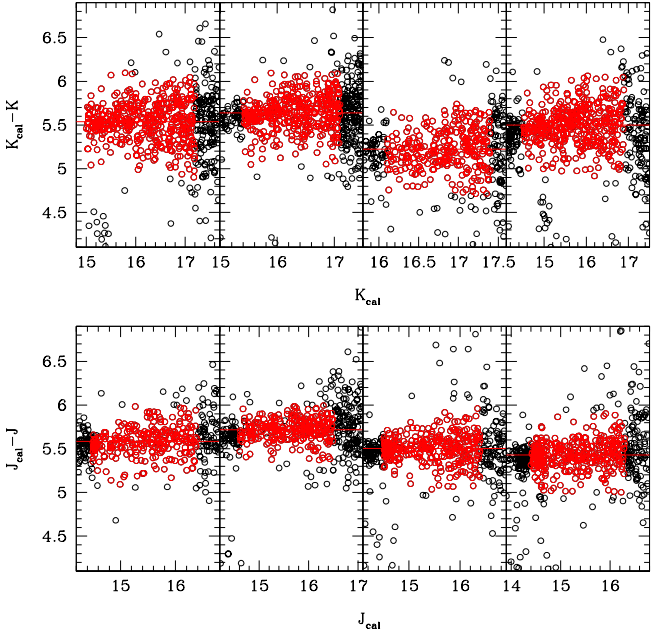}
        \caption{\small Calibration zero-points computed from the identification of stars in common between
        the HAWK-I and GeMS catalogues. Upper panels shows the zero-points for the four chips in the Ks-band,
        while lower panels show the same for the J-band. Stars surviving the selection criteria are plotted as
        red circles.}\label{cal_gems}
\end{figure}

We were able to exploit the HAWK-I spatial resolution to find about $1200$ stars
in common with each single-chip GeMS catalogue. We used these stars for the absolute photometric calibration by
applying the zero points shown in Fig.\ref{cal_gems}, computed according to a 3$\sigma$-clipping
iterative procedure in a range of magnitude suitably chosen to exclude faint stars and sources
that reached non-linearity in the long exposures (stars surviving this selection are plotted as
red circles in Fig.\ref{cal_gems}). We chose to apply constant zero points because
when looking for the best-fitting linear relations, the slopes turned out to be zero within the uncertainties
of the fit for all the 8 chips.
Similarly to what was found by \cite{turri}, in the K-band one of the chips appears 
to have an efficiency $\sim0.2$ mag higher than the others.
Once calibrated, the four catalogues have been merged together to create the final GeMS catalogue. 
The errors on the inter-chip calibration has been computed as the sum in quadrature of the standard deviations around 
the mean zero point of each chip, and are $\sigma_{cal,J}=0.02$ mag and 
$\sigma_{cal,K}=0.02$ mag.

The calibration procedure also allowed us to verify that the adopted cubically varying PSF model well describes 
the actual PSF variation across GeMS FoV. In fact, by computing the zero-points in smaller sub-fields, we found that 
they all agree within a 1 $\sigma_{cal}$ uncertainty with those adopted for the calibration. The only exceptions were 
found for the upper-right corner of chip 4 and the lower-right corner of chip 1, where the K-band zero-points differed 
by 2.5 $\sigma_{cal,K}$ and 2 $\sigma_{cal,K}$, respectively. Indeed these sub-fields correspond to the regions with 
the largest measured FWHM (see the left panel of Fig.\ref{fwhm}). However, as will be specified in the next Section, 
these two chips will not be used for the age estimation, and therefore they will not affect our final results.

\section{Colour magnitude diagrams}\label{cmd}

\subsection{IR CMD}

\begin{figure}
    \includegraphics[width=\columnwidth]{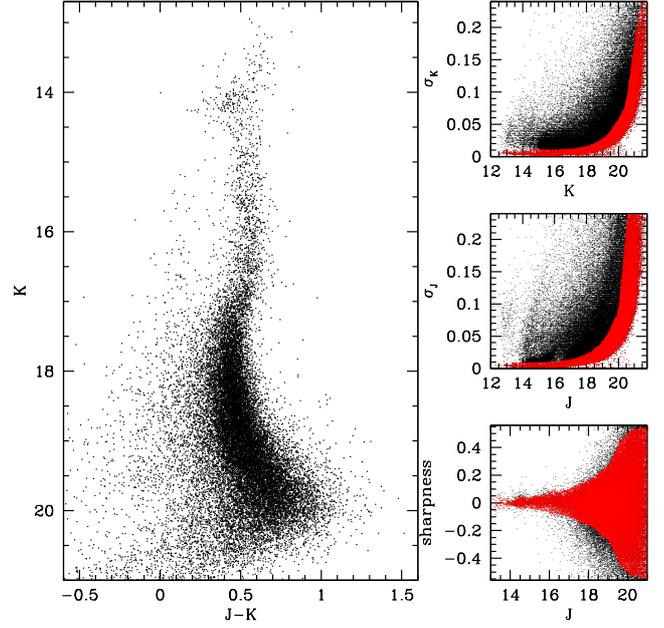}
        \caption{(K, J-K) CMD of NGC~2808 from our GeMS photometry. Stars shown in the CMD have been
        selected according to their photometric errors and sharpness (red dots in the right panels) and including
        only sources located farther than $30$\arcsec from the cluster centre.}\label{kjk_gems}
\end{figure}

The (K, J-K) CMD obtained from the final GeMS catalogue is shown in Fig.\ref{kjk_gems}. Stars have been selected according to
their photometric error and sharpness, as shown in the right panels of the Figure, and their distance
from the centre of the cluster (\citealt{goldsbury10}), excluding those within a radius of 30\arcsec~ to avoid the severe crowding
affecting those regions.

\begin{figure}
    \includegraphics[width=\columnwidth]{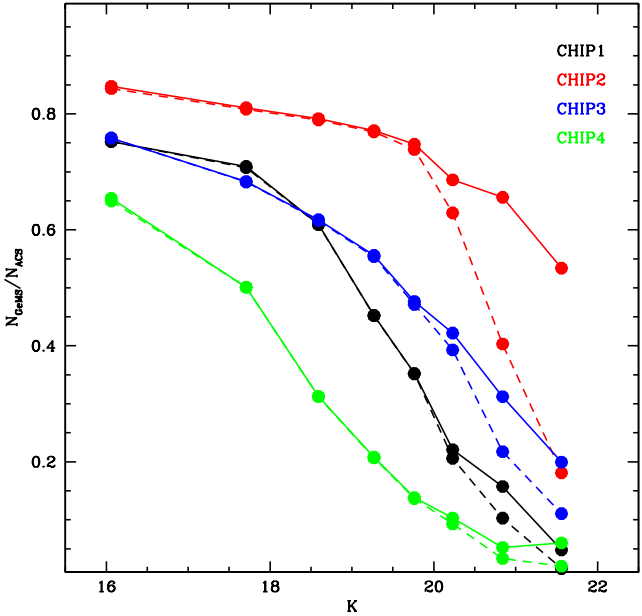}
        \caption{Completeness of our GeMS photometry relative to HST from the lower RGB down to the MS. Solid lines describe the trend
        for the K-band, dashed lines for the J-band. Each chip is shown with a different colour, and the
        best performing chip in both filters turned out to be chip2.}\label{compl}
\end{figure}

The CMD shows the evolutionary sequences from the MS to the red giant branch (RGB), where
the short exposures reach non-linearity (at $K\sim14.5$) and then saturate before the RGB-tip.
Moving towards fainter stars, the CMD reaches several magnitudes below the MSTO, but it is not 
deep enough to reach the MS-knee. 
The not ideal observing conditions might have affected the depth of our photometry, especially
in the J-band. To quantify how the completeness of our photometry has suffered for this effect,
we cross-matched our GeMS catalogue with the optical HST photometry
coming from the ACS survey of Galactic GCs (\citealt{sarajedini07}). The comparison between the number
of stars detected in the two catalogues in different bins of magnitude gives their relative completeness.
Fig.\ref{compl} shows such a relative completeness for each chip of GeMS as a function
of the observed GeMS K magnitude, from the lower RGB to the faint MS. The K magnitude in the X-axis has been obtained
from the ACS V magnitudes using the (K, V-K) ridge-line for the conversion. 
The trends for the relative completeness of the GeMS K-band photometry is shown with
solid lines, while those of GeMS J-band photometry with dashed-lines. As expected V-band ACS deep photometry has a larger completeness than that 
reached with K or J bands GeMS photometry. In particular in K-band there is only one chip that maintains a relative completeness 
above the $50$\% level (i.e. chip 2, red line in Fig.\ref{compl}). In J-band
the relative completeness drops below $50$\% at quite brighter magnitudes, with the best case being again chip 2 at V$\simeq24$.
Therefore, as already observed in \cite{turri}, the optical-IR K,V-K CMD is expected
to be much deeper, and getting rid of the limiting effect of J magnitudes it will reach the MS-knee
allowing us to obtain an accurate estimate of the cluster age.

\subsection{Optical-IR CMD}

\begin{figure}
    \includegraphics[width=\columnwidth]{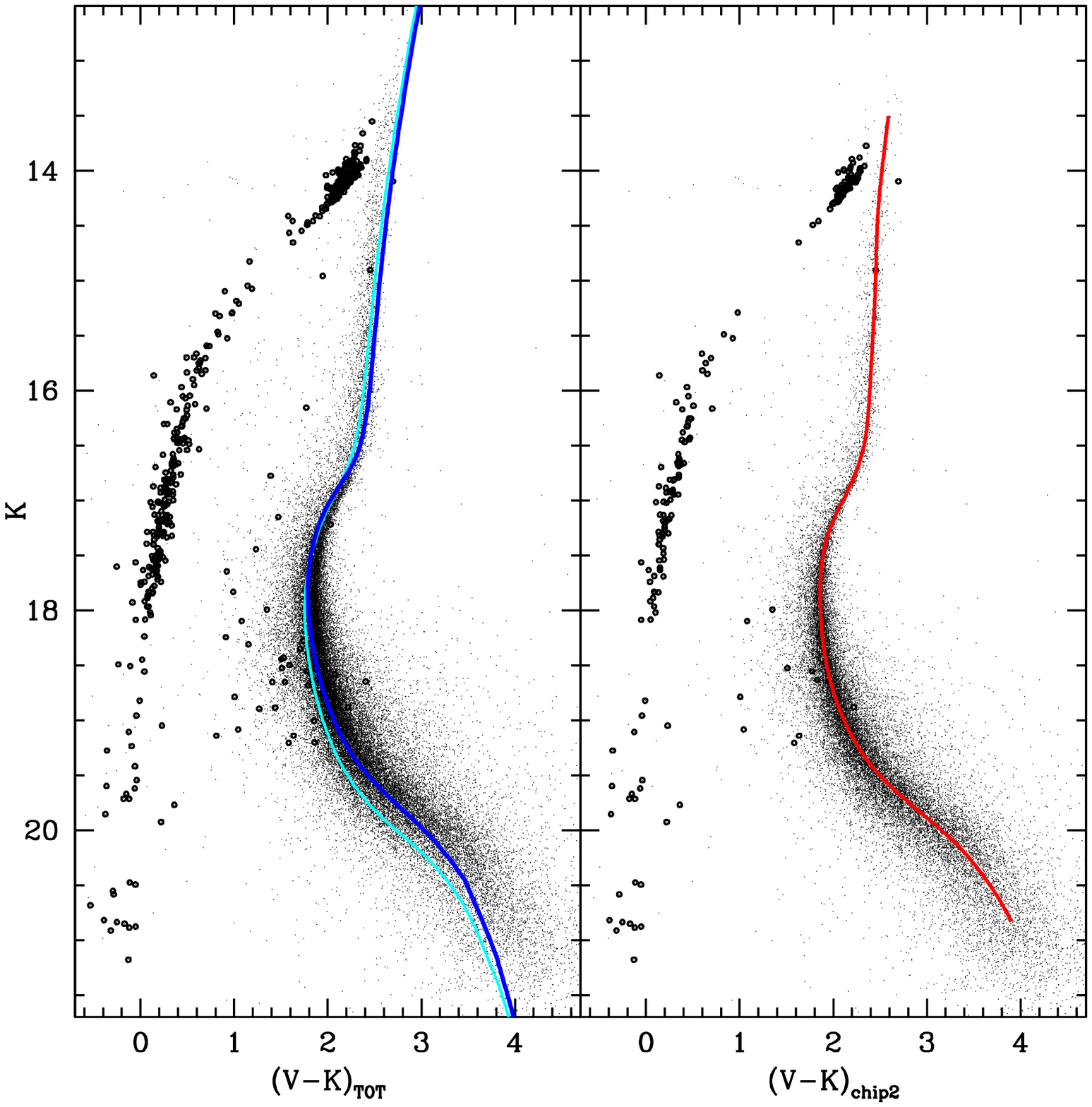}
        \caption{\small (K, V-K) CMD of NGC~2808 obtained from the combination of our GeMS photometry and the HST/ACS catalogue
        of \citep{sarajedini07}. The left panel shows the CMD of the entire GeMS sample, fit by a theoretical model taken from
        \citep{vdb12}. An Y-enhanced model (Y$=0.33$) is also plotted in cyan for sake of comparison. 
        The right panel shows only stars belonging to the best chip 2 and the corresponding ridge line is superimposed in red. 
        HB stars are highlighted with thick symbols.}\label{kvk}
\end{figure}

Left panel of Fig.\ref{kvk} shows the K,V-K CMD obtained for the stars in common between the ACS
catalogue and our GeMS photometry, cleaned for photometric errors and sharpness. The optical magnitudes
are calibrated to the Johnson-Cousins system (they correspond to the V$_{ground}$ and I$_{ground}$ of the publicly
available catalogue). 

Data Plotted in this figure display several interesting features worth being discussed in more detail. \\
($i$) It is immediately clear that the limitating effect of the J band has been removed and the optical-NIR CMD is much deeper then the purely IR one, 
spanning from the RGB bump  (K$\sim$13.5 mag, see also Fig.\ref{cal_hawki}) down to the lower main sequence (K$\sim21.5$ mag). \\
($ii$) The current photometry covers the entire range of HB stars, namely from K$\sim14$ mag,
typical of red HB stars, down to K$\sim21$ mag, typical of extreme HB stars. HB stars were selected from the optical
CMD of the ACS catalogue, and marked with filled circles in Fig.\ref{kvk}. Dating back to the seminal investigations by 
\cite{ferraro98} and by \cite{castellani06} the HB morphology of NGC~2808 has been a longstanding puzzle, since it is the 
so-called prototype of multi-modal HB morphology. When moving from cooler to hotter stellar structures, the cluster HB shows 
sizeable samples of red and blue HB stars and a very limited sample of RR Lyrae stars (\citealt{kunder13b}). Its luminosity 
function shows two well defined gaps and a tail of stellar structures approaching the typical range of sub-dwarf B- and sub-dwarf O- stars 
(\citealt{castellani07, moehler11, latour14}). The comparison with the HB luminosity function of Galactic globulars showing similar 
extended HB morphologies has been hampered by the use of either optical or near UV colours that have limited sensitivity when 
moving from extreme HB (T$_{eff}$=35,000 K) to red HB (T$_{eff}$=5,000 K) stars. The key advantage in using the 
V-K colour is that it is a solid temperature indicator for giant stars (\citealt{dibenedetto05}) and the quoted temperature range is 
covered by roughly three magnitudes in V-K. Current optical-NIR photometry (see e.g. \citealt{libralato14}) are paving the road 
for constraining the secondary features (gaps; jumps, see \citealt{grundahl99}) of the HB luminosity function. Moreover, they are 
also playing a fundamental role in using hot HB stars as stellar tracers (\citealt{kinman12}).\\  
($iii$) There is evidence that a few hot HB candidates attain, at fixed K-band magnitudes, colours that are systematically cooler 
(V-K=0.2-0.3 mag) than typical extreme HB stars. Similar objects were called by \cite{castellani06} HB peculiar, suggesting 
that they might be intrinsic binary candidates. The optical-NIR colours seem to support this working hypothesis.   

In the left panel of Fig.\ref{kvk} we show a comparison between the photometry and a theoretical model interpolated from the set of isochrones provided by \cite{vdb12}, 
suitably selected according to the age (11 Gyr, \citealt{vdb13}), metal composition ([Fe/H]$=-1.13$ dex, [$\alpha$/Fe]$=+0.3$ \citealt{carretta15}) 
and using a standard helium content (Y$=0.25$). We placed the isochrone onto the observed CMD 
and found that the best fit was achieved by assuming a colour excess E(B-V)=0.17 mag and a
distance modulus of m-M$_{0}=15.00$ mag. These values are in agreement with previous estimates available
in literature, such as \citealt{piotto07, kunder13b}, who found E(B-V)$=0.18$ mag, m-M$_{0}=15.0$ mag and E(B-V)$=0.17$ mag, 
m-M$_{0}=15.04$ mag, respectively.
From such a comparison, it is immediately clear that this colour combination goes much deeper than the purely near-IR, easily reaching the faint MS knee.
Since its location in magnitude and colour
is essentially independent of cluster age at fixed chemical composition, it has been first used by
\cite{bono10c} and then in other works (\citealt{dicecco15, monelli15}) to determine the absolute age of GCs
with an accuracy that is typically two times better than obtained by other age indicators. 
In fact, the use of a relative difference between the magnitude of the MSTO and the MSK suppresses the
effect that uncertainties on several parameters (such as distance and reddening, see \citealt{monelli15} for 
more details) have on the accuracy of the age estimate. Moreover,
the MSK is located in a region of the MS that is only marginally affected by uncertainties on the
treatment of the convection in theoretical models, since these convective motions are nearly
adiabatic (\citealt{saumon08}).

For these reasons and with the aim of determining the most accurate age estimate to date for NGC~2808, we computed the mean 
ridge-line of the K, V-K CMD following the method described in details in \cite{dicecco15} and using only the stars detected
in the GeMS chip with the best AO-correction efficiency and stability (chip2, see Fig.\ref{fwhm}), 
and therefore deeper and more complete photometry (see Fig.\ref{compl}). This choice allows us to avoid systematic effects 
introduced by the calibration procedure.
The adopted method consists of a numerical algorithm that pin points the peaks of the iso-contours described by the 
K, V-K CMD to provide the ridge-line first guess. This solution is then fit with a bi-cubic spline and smoothed\footnote{A 
graphic representation of these steps is presented in Fig.2 of \cite{dicecco15}.}. Finally, in order to be as accurate 
as possible in the determination of the relevant points (MSTO and MSK), we re-sampled the ridge-line by steps of 0.001 mag.
The CMD and ridge line are shown in the right hand panel of Fig.\ref{kvk}.

\section{Results}\label{results}

Several previous estimates of the relative age of NGC~2808 in the literature agree that this cluster has a younger age than
the average of Galactic GCs with a similar metallicity.
By using the so-called vertical method measuring the relative magnitude difference between the ZAHB and the MSTO, 
\cite{deangeli05} found NGC~2808 to be about the 23\% younger than clusters of similar metallicity, that is $\sim8.4\pm0.9$ Gyr old.
Following the MS-fitting method and comparing TO luminosities, \cite{marin09} estimated a difference of about 15\%, corresponding to
an age of $10.9\pm0.4$ Gyr, and classified NGC~2808 as belonging to their class of young clusters. 
By studying the peculiar morphology of the HB, \cite{dotter10} and \cite{milone15} similarly found a 10-15\% younger relative age.
Using an improved version of the vertical method, which takes into account constraints coming from the HST CMD fitting,
\cite{vdb13} also estimated a younger age, and anchoring it to an absolute scale they quoted an age for NGC~2808 of $11.0\pm0.4$ Gyr.
All these relative age measurements have small associated uncertainties, but we underline that when translating them to absolute ages
an additional term of $1-2$ Gyr due to uncertainties in reddening, distance and photometric zero points must be taken into account
(\citealt{monelli15}).
On the other hand, \cite{piotto07} found that the best fit of their triple MS was achieved using a 12.5 Gyr isochrone.

\begin{figure}
    \includegraphics[width=\columnwidth]{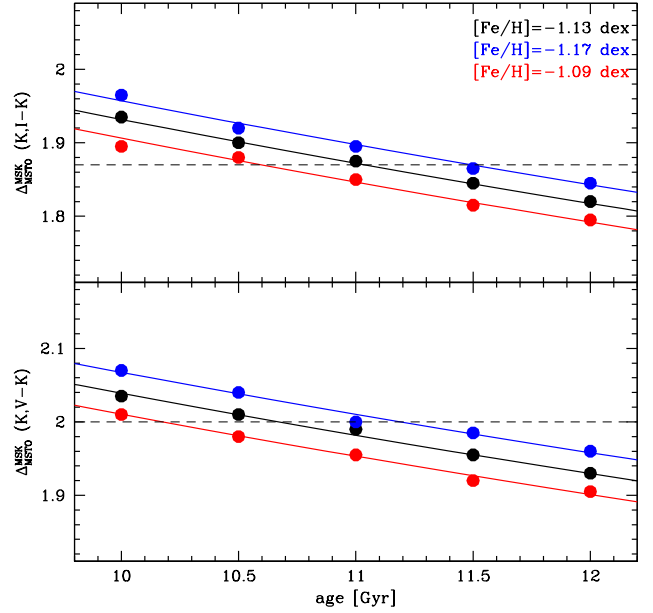}
        \caption{\small {\it Upper panel:} theoretical relation between $\Delta_{MSK}^{MSTO}$ and cluster age in the K,I-K plane. The different
        lines correspond to metallicity values differing by 0.04 dex (see the colour-coded labels). The dashed black line correspond to the observed value.
         {\it Lower panel:} same but for the K,V-K plane.}\label{rel_delta}
\end{figure}

Given this picture is in overall agreement, but with a few contradictory findings,
our work aims to resolve the age issue by providing a more accurate determination of the absolute age for this cluster.
Firstly, we selected the theoretical models of \cite{vdb12}, which fit very well the MS in the observed CMD shown in Fig.\ref{kvk}, 
and re-sampled them to the same magnitude step as that used for the ridge-line. Then,
we derived linear relations between the $\Delta_{MSTO}^{MSK}$ parameter and cluster age and metallicity. We built relations for
all the combinations of filters with K able to reach the MSK (to minimise
the possible effect of differential reddening), that is the K,V-K and the K,I-K planes.
The result of this bi-parametric fit is shown as lines in Fig.\ref{rel_delta}, where the black line corresponds to
the relation obtained for [Fe/H]$=-1.13$ dex, and the red and blue lines correspond to the fit for a difference in [Fe/H] of $\pm0.04$,
respectively ($\sigma_{[Fe/H]}=0.04$ is taken from \citealt{carretta15}). The analytic form of the two fits, together with the 
uncertainties on each coefficient, are
\begin{eqnarray}
 \log (age) &=& (1.893\pm0.060)-(0.723\pm0.023)*\Delta_{MSK}^{MSTO} - \nonumber\\ 
 &&(0.516\pm0.031)*[Fe/H]
\end{eqnarray}
for the K,V-K plane and
\begin{eqnarray}
 \log (age) &=& (1.839\pm0.063)-(0.692\pm0.024)*\Delta_{MSK}^{MSTO} - \nonumber\\
 &&(0.442\pm0.033)*[Fe/H]
\end{eqnarray}
for the K,I-K plane.

As a further check, we also computed the relation between the absolute K-magnitude of the MSTO (M$_{K,MSTO}$) and the cluster age and metallicity.
This relation will lead to a less accurate age determination, but will be used to verify whether an overall
agreement between the two methods is found.
The linear best fit we found is:
\begin{eqnarray}
 \log (age) &=& (-0.919\pm0.051)+ (0.661\pm0.016)*M_{K,MSTO} - \nonumber\\
 &&(0.074\pm0.021)*[Fe/H]
\end{eqnarray}

As a second step, we used the best-fitting ridge-lines (for the K,V-K CMD it is shown in the right panel of Fig.\ref{kvk})
to determine the observed $\Delta_{MSTO}^{MSK}$ and K$_{MSTO}$. The former values were measured as described in detail in \cite{dicecco15},
and are $\Delta_{MSTO}^{MSK}=2.00$ mag in the K,V-K plane and $\Delta_{MSTO}^{MSK}=1.87$ mag in the K,I-K plane 
(see the dashed horizontal lines in Fig.\ref{rel_delta}). The K$_{MSTO}$ value is the bluest point of the ridge-line,
and was found to be M$_{K,MSTO}=2.84$ mag. 
By adopting these values in the above equations, the absolute age of NGC~2808 as measured from the difference between the 
magnitude of MSTO and MSK is t$=10.7$ Gyr for the K,V-K case and t$=11.1$ Gyr for K,I-K. The careful estimate of the overall uncertainties 
is addressed in the following section. As a sanity check, we also computed the absolute age as measured from the magnitude of the MSTO.
The value of t$=11.0$ Gyr is in good agreement with the previous findings. The good fit provided by the $11$ Gyr old isochrone
already described in Fig.\ref{kvk} is a further proof of the validity of all our measurements.

\subsection{Age uncertainties}

To accurately compute uncertainties on the absolute ages, we have to carefully take into account
the possible sources of errors affecting both the photometry and the comparison with the theoretical models.
We start with the uncertainties affecting the position of both the MSK and the MSTO.

($i$) {\it MSK and MSTO determination}: the MSK is defined as the point with the minimum curvature of the low 
MS ridge-line. To compute the curvature parameter, we took each point of the ridge-line and looked for the circumference 
that connects it with the two points of the ridge-line located at $\pm0.5$ mag. The circumference with the largest radius 
(i.e. the minimum curvature) defines the MSK. The source of uncertainty introduced by this method is related the size of the 
magnitude step used to select the three circumference points. To quantify it, we re-computed the location of the MSK by 
varying such step from 0.1 mag to 0.8 mag (the limit at which the algorithm is still able to detect the MSK, since for larger 
steps it ceases to find the three points before reaching the MSK itself). Within this interval, the magnitude of the MSK varies 
by $\pm0.013$ mag. We therefore adopt this value as MSK determination uncertainty.  
On the other hand, the MSTO is defined as the bluest point of the ridge-line. In this case, it turned out that the bluest colour 
value is shared among $10$ adjacent magnitude sampling of the ridge-line (each of 0.001 mag size). For this reason, we adopted 
their average value as magnitude of the MSTO and we assume as uncertainty on its determination $\pm0.005$ mag.

($ii$) {\it Photometric uncertainty:} at the level of the TO, the average uncertainties for the stars in our final catalogue
are of the order of 0.008 mag, while at the faint MSK this increases to 0.035 mag. The sum in quadrature gives
a total internal error on $\Delta_{MSTO}^{MSK}$ of 0.036 mag. We underline that this is the only contribution from the photometry
to the error on $\Delta_{MSTO}^{MSK}$.

($iii$) {\it Intra-chip calibration uncertainty:} to avoid this term we only use the stars on the best chip of the GeMS dataset.

Now we list the uncertainties affecting the position of the MSTO only.

($i$) {\it Absolute calibration uncertainty:} this only affects the MSTO, and as pointed out in Sect.\ref{data}
it amounts to 0.02 mag.

($ii$) {\it Differential reddening:} this would affect both the MSTO and MSK magnitudes if they 
were not homogeneously distributed across the FoV. However, the choice of the most photometrically complete chip ensures such uniformity
and thus this term affects only the MSTO. The use of K-band minimises its
contribution. According to \cite{bedin00} NGC~2808 suffers only a  small amount of differential reddening 
($\delta$E(B-V)$\simeq0.02$) which when multiplied by the extinction coefficient of the K-filter (A$_{K}=0.35$, \citealt{cardelli89})
gives a tiny contribution of 0.007 mag.

($iii$) {\it Distance uncertainty}: \cite{kunder13b} estimated an uncertainty on their RR Lyrae variable stars distance modulus
(m-M$_{0}=15.04$) of 0.13 mag. We adopt this value as distance uncertainty, underlining that it affects only the MSTO.

($iv$) {\it Absolute colour-excess uncertainty}: the value adopted in our isochrone fitting E(B-V)$=0.17$ mag is in good agreement
with previous works. \cite{bedin00} quoted E(B-V)=$0.19$ mag, which also has been used by \cite{piotto07,
milone15}. \cite{kunder13b} used E(B-V)$=0.17$ mag, while \cite{harris96} adopts E(B-V)$=0.22$ mag. Given this overall agreement,
we adopt $0.05$ mag as a conservative uncertainty. Also in this case, only the MSTO is affected.

All the quoted terms contribute to the uncertainty on the age from the photometry. 
By summing together these contributions, we get $\sigma_{\Delta}=0.038$ mag and $\sigma_{MSTO}=0.145$ mag.
This translates into age uncertainties of $\sigma_{t,\Delta}=0.7$ Gyr and $\sigma_{t,MSTO}=2.7$ Gyr. In addition, we have to take into account
the effect of metallicity and helium content.

By following the relations in the previous section, an error of $\sigma_{[Fe/H]}=0.04$
(\citealt{carretta15}) is converted to $\sigma_{t,\Delta}=0.45$ Gyr and $\sigma_{t,MSTO}=0.1$ Gyr.
The contribution to the error budget coming from helium-enhancement is not possible for us to accurately estimate as our photometry is not good enough, and the combination
of filters not well suited\footnote{UV filters are much more efficient in enhancing the effect of different
chemical composition on the photometry of GC multiple generations, see e.g. \cite{milone15}}, to separate the populations of NGC~2808 with different Y abundance. For this reason, it is not trivial to
quantify the possible effect of an Y-enhanced population on our age estimate.
By using the set of theoretical models of \cite{vdb12}, we found that in comparison with a standard helium model, an isochrone with Y$=0.33$
(and same metallicity and age) is 0.17 mag fainter for M$_{K,MSTO}$ and 0.24 mag fainter for M$_{K,MSK}$ (see the cyan line in the left-hand panel of
Fig.\ref{kvk}). The difference in $\Delta_{MSTO}^{MSK}$ would
therefore be 0.07 mag larger. This means that any contamination from a Y-enhanced population will artificially increase $\Delta_{MSTO}^{MSK}$, and thus 
we would over estimate the age of the cluster. A Y$=0.33$ population
with the same parameters would appear 1.3 Gyr older. However, \cite{piotto07} showed that the total fraction of Y-enhanced stars
is $\sim28$\%. Moreover, the Y-rich populations have been found to be located preferentially in the 
innermost region of the cluster, which we excluded to avoid the effects of crowding. Unfortunately,
no quantitative determinations of the radial distribution of the multiple sequences in NGC~2808 is available in the literature. Therefore, we can only
make an estimate of the possible contamination, with the limitation that it has to be smaller than 28\%. 
If we make the conservative assumption of a 20\% Y-rich contamination, then the
effect would be an increase of $\Delta_{MSTO}^{MSK}$ by 0.014 mag, corresponding to 0.25 Gyr.
The resulting effect on the age coming from the MSTO is of 0.05 Gyr.

Therefore, our final age and error estimates are t$_{\Delta,KVK}=10.7\pm0.7$ (intrinsic) $\pm0.45^{+0.25}_{-0.0}$ (metallicity and helium terms) Gyr, 
t$_{\Delta,KIK}=11.1\pm0.7\pm0.45^{+0.25}_{-0.0}$ Gyr and t$_{MSTO}=11.0\pm2.7\pm0.1^{+0.05}_{-0.0}$ Gyr. 
The ages determined by the $\Delta_{MSTO}^{MSK}$ method are more than twice as accurate as that coming from the MSTO method.
Because there is no reason to prefer the age coming from a particular filter combination, we decide to average
the two values coming from $\Delta_{MSTO}^{MSK}$. In this way we obtain the final and most accurate absolute age obtained for NGC~2808 
which is t$_{\Delta}=10.9\pm0.7\pm0.45^{+0.25}_{-0.0}$. This value is compared to the others available in literature in Table \ref{ages}.

We underline that this age estimate is directly comparable to all those achieved via the $\Delta_{MSTO}^{MSK}$ method in the near-IR so far, since they are all based 
on relations built using the theoretical models of \cite{vdb12}. A detailed study on the effects on the age measurements coming from the 
use of different set of isochrones will be presented in a forthcoming paper (Fiorentino et al. in prep.).


\begin{table}
\caption{List of NGC2808 age estimates.}             
\label{ages}      
\centering                          
\begin{tabular}{c c c }        
\hline\hline                 
Method & Reference & Age (Gyr) \\    
\hline                        
                                                &   &                               \\
   $\Delta_{MSTO}^{MSK}$                        & 1 & $10.9\pm0.7\pm0.45^{+0.25}_{-0.00}$   \\      
                                                &   &                               \\
   M$_{K,MSTO}$                                 & 1 & $11.0\pm2.7\pm0.10^{+0.05}_{-0.00}$   \\
                                                &   &                               \\
   $\Delta V_{ZAHB}^{MSTO}$                     & 2 & $8.4\pm0.9^{\ast}$                   \\
                                                &   &                               \\
   Isochrone fitting                            & 3 & 12.5                          \\
                                                &   &                               \\
   Isochrone fitting                            & 4 & $10.9\pm0.4^{\ast}$                  \\
                                                &   &                               \\
   $\Delta V_{ZAHB}^{MSTO} +$ Isochrone fitting & 5 & $11.0\pm0.4^{\ast}$                  \\
                                                &   &                               \\
\hline                                   
\end{tabular}
\tablefoot{$^{\ast}$ Errors on the {\it relative} age.}
\tablebib{(1)~This work;
(2) \citet{deangeli05}; (3) \citet{piotto07}; (4) \citet{marin09};
(5) \citet{vdb13}.
}
\end{table}

\section{Conclusions}\label{concl}

In this paper, we presented the most accurate determination of the absolute age of the Galactic GC NGC~2808 using the MSK method
described by \cite{bono10c}. To reach such a faint feature of the CMD, we analysed deep GeMS IR images and combined
the photometry in the K-band with V-band magnitudes coming from HST.
The resulting CMD did not allow us to clearly separate the multiple sequences
as has been done using HST photometry in the UV (\citealt{milone15}), but provides for the first time K-band magnitudes
of extreme-HB stars in GCs.

By comparing our measured difference between the magnitude of the MSTO and that of the MSK in the K,V-K and K,I-K planes
with the output of theoretical model from \cite{vdb12}, we found an age of t$=10.9$ Gyr.
An accurate analysis of the error budget shows that the overall uncertainty due to the photometry
is $\pm0.7$ Gyr. An additional term due to the uncertainty
on the cluster metallicity is $0.45$ Gyr, and another due to possible contamination
from Y-enhanced stars could make the cluster as much as $0.25$ Gyr older.

Even considering all the sources of errors, our absolute age estimate is the most accurate obtained
for this cluster, and agrees well with the previous studies that found NGC~2808 
to be a cluster slightly younger than other clusters of similar metallicity.
This analysis seems to rule out the estimate of $8.4$ Gyr obtained in \cite{deangeli05} through the vertical method
and the value of $12.5$ Gyr used by \cite{piotto07} to fit the three sequences in their CMD.

Accurate measurements of relative ages have led to the possible discovery of a bimodal age-metallicity relation for Galactic GCs 
(\citealt{marin09, leaman13}). Such a bimodality is interpreted in terms of a different origin, where the younger clusters may have been 
lost by bigger systems accreted by the Milky Way, like dwarf galaxies. If this hypothesis is correct, then very accurate absolute ages, 
coupled with complete chemical screenings and orbit determinations, are required in order to be able to link these clusters to their 
possible progenitor. Future GCs age measurements with the method used in this work will provide the homogeneous set of estimates 
required to address this standing problem.

\begin{acknowledgements}

We thank the anonymous Referee for his/her suggestions which helped us to significantly improve the presentation of our results.  
Based on observations obtained at the Gemini Observatory. Acquired through the Gemini Science Archive and processed using the Gemini IRAF package.
This publication makes use of data products from the Two Micron All Sky Survey, which is a joint project of the University of 
Massachusetts and the Infrared Processing and Analysis Center-California Institute of Technology, funded by the
National Aeronautics and Space Administration and the National Science Foundation.
GF and DM has been supported by the FIRB 2013 (MIUR grant RBFR13J716).

\end{acknowledgements}

%
   \bibliographystyle{aa} 
   \bibliography{ms.bib} 
%




\end{document}